\documentclass[aps,prd,twocolumn,groupedaddress]{revtex4-1}

\usepackage{amsmath,amssymb,bm}
\usepackage{graphicx}
\usepackage{epstopdf}
\usepackage{amsfonts}
\usepackage{amssymb}
\usepackage{amsbsy}
\usepackage{amsmath}
\usepackage{latexsym}
\usepackage{sansmath}
\usepackage{mathrsfs}
\usepackage{bm}
\usepackage{color}
\usepackage{comment}
\usepackage{csquotes}								
\usepackage{physics}
\usepackage{eufrak}
\usepackage{accents}
\usepackage[colorlinks=true,breaklinks]{hyperref}
\usepackage[caption=false]{subfig}
\usepackage[utf8]{inputenc}

\providecommand{\ham}{\mathcal{H}}
\newcommand{\M}{M_\text{Pl}}
\newcommand{\altdot}[1]{\accentset{\circ}{#1}}
\newcommand{\di}{\partial}
\newcommand{\eff}{\text{eff}}
\newcommand{\phys}{\text{phys}}
\providecommand{\Msig}{M_\sigma}

\newcommand{\eq}[1]{Eq.\,(\ref{#1})}

\def\be{\begin{equation}}
\def\ee{\end{equation}}
\def\bea {\begin{eqnarray}}
\def\eea {\end{eqnarray}}
\def\nn {\nonumber}
\def \p {\partial}
\def \pp {\mathsf{p}}
\def \l {\left}
\def \r {\right}

\begin{document}
\title{Natural inflation from polymer quantization}

\author{Masooma Ali} \email{masooma.ali@unb.ca} \affiliation{Department of Mathematics and Statistics, University of New Brunswick, Fredericton, NB, Canada E3B 5A3}

\author{Sanjeev S.\ Seahra} \email{sseahra@unb.ca} \affiliation{Department of Mathematics and Statistics, University of New Brunswick, Fredericton, NB, Canada E3B 5A3}

\begin{abstract}

We study the polymer quantization of a homogeneous massive scalar field in the early universe using a prescription inequivalent to those previously appearing in the literature.  Specifically, we assume a Hilbert space for which the scalar field momentum is well defined but its amplitude is not. This is closer in spirit to the quantization scheme of loop quantum gravity, in which no unique configuration operator exists. We show that in the semi-classical approximation, the main effect of this polymer quantization scheme is to compactify the phase space of chaotic inflation in the field amplitude direction.  This gives rise to an effective scalar potential closely resembling that of hybrid natural inflation.  Unlike polymer schemes in which the scalar field amplitude is well-defined, the semi-classical dynamics involves a past cosmological singularity; i.e., this approach does not mitigate the big bang.

\end{abstract}

\maketitle

\section{Introduction}

Although proposed by Guth \cite{Guth:1981} to explain the flatness of the universe and non-observation of magnetic monopoles, the inflationary epoch in the standard model of cosmology provides a mechanism to solve the horizon and entropy problems, as well as to generate the primordial perturbations that seed the observed large scale structure in the universe. Many models for inflation have been proposed, several of which involve scalar fields coupled to gravity. A large class of these models rely on the slow roll approximation, where an inflaton field olls to the minimum of its potential while undergoing negligible acceleration \cite{Steinhardt:1984}.  However, conventional single field slow roll inflation suffers from a fine tuning problem. In order to provide sufficient inflation to solve the horizon problem and result in an amplitude of density fluctuations consistent with observations, the inflaton potential is required to be very flat. Specifically, the ratio of the height of the potential to the fourth power of its width must be less than $10^{-6}$ \cite{Adams:1991}.  

The natural inflation model proposed by Freese et al \cite{Freese:1990} addresses this by employing a shift symmetry to protect the flatness of the potential. The inflaton is modelled as a pseudo Nambu Goldstone boson produced due to spontaneous symmetry breaking of a global shift symmetry (like an axion) that subsequently undergoes slow roll due to explicit symmetry breaking. The simplest potential for natural inflation is given by
\be
V_\text{NI}(\phi)  = \Lambda^4 \l[ 1 - \cos(\frac{\phi}{ f }) \r] = 2\Lambda^{4} \sin^{2} \l(\frac{\phi}{ 2f }\r).  \label{NI}
\ee
The two mass scales $\Lambda$ and $f$ associated with the explicit and spontaneous symmetry breaking. In order to produce sufficient inflation and match the observed scalar spectral index one requires $\Lambda  \sim 10^{16}$ GeV and $f \gtrsim 10^{19}$ GeV \cite{Freese:2015}. This implies a tensor-scalar ratio greater than favoured by Planck data \cite{Planck}.  Since the initial model appeared, other theoretical mechanisms have been proposed to obtain a potential similar to $V_\text{NI}$ that are consistent with observation.  For example, some authors have discussed the possibility that natural inflation is a consequence of the weak gravity conjecture \cite{delaFuente:2014}.  The hybrid natural inflation (HNI) model \cite{Ross:2009hg,Ross:2010fg,Carrillo-Gonzalez:2014tia,Vazquez:2014uca,German:2016umn,Ross:2016hyb,German:2017pfu} introduces a new parameter $\rho_\text{v}$ in the potential that acts as a cosmological constant:
\be
\label{eq:HNI Potential}
V_\text{HNI}(\phi) = \Lambda^4 \l[ 1 - \cos \l(\frac{\phi}{f} \r) \r] + \rho_\text{v}.
\ee
This model relies on a second scalar field to end inflation (the so-called ``waterfall field''), and allows for observationally sufficient accelerated expansion even with sub-Planckian symmetry breaking for the Goldstone field.  It is therefore consistent with Planck data for a wide range of parameters \cite{Ross:2016hyb,German:2017pfu}.  

In this paper, we show that a HNI type of potential with two mass scales can also be obtained from a polymer quantized minimally coupled scalar field propagating on a Friedmann-Robertson-Walker (FRW) background. Polymer quantization is a background independent quantization scheme employed in loop quantum cosmology (LQC), where the holonomy-flux algebra of loop quantum gravity (LQG) in a minisuperspace setting is represented on the space of square integrable functions on the Bohr compactification of the real line. Polymer quantization is a general scheme that can be used to quantize any classical field. Unlike LQC, where the geometry is polymer quantized, we study a polymer quantized scalar field on a fixed background. Polymer quantized matter has been studied by several authors \cite{Ashtekar:2002,Hossain:2009vd,Hossain:2009ru,husain:2010,Hossain:2010eb,Hossain:2010wy,Seahra:2012un,Husain:2013zda,Hassan:2014sja,Kajuri:2015,Hassan:2017cje}. The polymer quantization of the free scalar field in an isotropic and homogenous setting was studied in \cite{Hossain:2009ru}, while the minimally coupled scalar field was studied in \cite{Hassan:2014sja}. The effective semiclassical dynamics detailed in \cite{Hassan:2014sja} indicates that polymer quantized scalar fields with quadratic potentials generically result in an early time ``polymer" inflation phase followed by a slow roll inflation phase.   The early time polymer phase closely resembles past eternal de Sitter inflation; hence, such models effectively tame the big bang singularity of classical general relativity.

Below, we polymer quantize the same classical dynamics as in \cite{Hassan:2014sja} albeit with a different (inequivalent) prescription for quantization. The polymer quantization we use is in the spirit of LQC, where the momentum operator is diagonal in some basis on the Hilbert space and no unique configuration operator exists (the version used in \cite{Hassan:2014sja} employs a diagonal configuration operator). A priori, there is no reason to choose one prescription over the other and our objective here is to show that each choice results in different semiclassical dynamics. The prescription used here results in a compact phase space for the semiclassical dynamics. (Compact phase spaces in cosmology have been discussed in other contexts \cite{Matos:2009hf,Mielczarek:2017zoq}).

This paper is organized as follows: In Section \ref{classical}, we present the classical theory of a scalar field minimally coupled to gravity under the assumptions of homogeneity and isotropy and Section \ref{polymer} gives details of the polymer quantization. In Section \ref{semiclassical} we study the effective semiclassical dynamics of the theory and compare it with observation.
\section{The Classical Theory} \label{classical}
In the Arnowitt-Deser-Misner (ADM) formalism of general relativity, the action of a scalar field minimally coupled to gravity is
\be
S = \int d^3x \, d\tau \l( \pi^{ab}\altdot{q}_{ab} + p_\phi \altdot{\phi} - \mathcal{NH} -\mathcal{N}^a \mathcal{C}_a \r),
\ee
with the Hamiltonian constraint $\ham$ and the diffeomorphism constraint $\mathcal{C}_{a}$ given by
\bea
\ham &=& \frac{1}{8\pi G\sqrt{q}} \l( \pi^{ab}\pi_{ab} - \frac{1}{2}\pi^2 \r) - 8\pi G \sqrt{q}^{(3)} R  \nn \\
 &+&  \l( \frac{p_\phi^2}{2\sqrt{q}} + \frac{1}{2}\sqrt{q}\p_a \phi \p^a \phi +\sqrt{q} V(\phi) \r), \nn\\ 
\mathcal{C}_a &=& -D_b \pi^b_a + p_\phi \p_a \phi. 
\eea
We use an open dot to represent the derivative with respect to the general time variable $\tau$; i.e., $\altdot{X} = dX/d\tau$.  We take the ``bare'' classical scalar field potential to be that of chaotic inflation:
\begin{equation}\label{eq:bare}
	V(\phi) = \frac{1}{2} m^2 \phi^{2}.
\end{equation}
The phase space variables are $(q_{ab},\pi^{ab}, \phi, p_\phi)$, and associated with the spatial metric $q_{ab}$ are the covariant derivative $D_a$ and the scalar curvature $^{(3)}R$.
In order to obtain the Friedmann-Robertson-Walker (FRW) solutions, homogeneity and isotropy are imposed at the level of the action. The reduced action for spatially flat FRW can be obtained using the parameterization
\be
q_{ab} = a^2(\tau) \delta_{ab}, \quad \pi^{ab} = \frac{p_a}{6a(\tau)}\delta^{ab},
\ee
where $\delta_{ab}$ is the $3$D flat Euclidean metric. The action is
\be
S_R =  \int d\tau \l( p_a \altdot{a} + p_\phi \altdot{\phi} - \mathcal{N}H_c \r),
\ee
where
\begin{align}
\nn H_c & = H_G + H _ \phi \\
& =  - \frac{p_a^2}{12 \M^2 \mathcal{V}_0 a} + \frac{\Pi_\phi^2}{a^3 \mathcal{V}_0} + a^3 \mathcal{V}_0 V(\phi). \label{ham_cnstrnt}
\end{align}
Here, $\M = 1 / 8 \pi G$ is the reduced Planck mass, $\mathcal{V}_0 = \int d^3x$ is the fiducial volume and we have redefined the conjugate momenta as the scalars 
\begin{equation}
	p_a  = -\frac{6\M^3 \mathcal{V}_0 a\altdot{a}}{\mathcal{N}},  \quad \Pi _ {\phi} = \frac{a^3 \mathcal{V}_0 \altdot{\phi}}{\mathcal{N}}.
\end{equation}
The relevant Poisson brackets are $\{a,p_a\} = 1$ and $\{\phi, \Pi_\phi \}=1$.

This reduced action is invariant under spatial dilations, a symmetry of spatially flat FRW. Under dilations we have
\begin{subequations}\label{eq:transformations}
\bea 
\vec{x} &\rightarrow& \ell \vec{x}, \label{eq:dilations}  \\
(a,p_a,\phi,p_\phi, \mathcal{V}_0) &\rightarrow& (\ell^{-1}a,\ell p_a, p_\phi,\phi,\ell^3 \mathcal{V}_0).
\eea
\end{subequations}
The reduced action is also invariant under time reparametrizations, i.e $H_c = 0$ is a first class constraint. Thus, the dynamics of the system are governed by Hamilton's equations and initial data satisfying $H_c = 0$. A reduction to  the physical degrees of freedom can be obtained by making a choice of time and solving the Hamiltonian constraint explicitly. Such a time gauge (choice of time) can be fixed by identifying some function of the phase space variables as the time parameter. Solving the Hamiltonian constraint for the conjugate variable to this time parameter then yields the physical Hamiltonian. A convenient choice is the ``e-fold'' time gauge:
\be
\tau = N \equiv \ln \l( \frac{a}{a_0} \r).
\ee
Here, $a_0$ is a constant determined by the choice of initial hypersurface. In order for our time parameter to be invariant under spatial dilations (\ref{eq:dilations}) we require the transformation $a_0 \rightarrow \ell^{-1} a_0$ under dilations. The physical Hamiltonian is proportional to the momentum conjugate to the time variable,
\be
p_N = ap_a =a_0 e^{N}p_a.
\ee
Thus, the physical Hamiltonian is 
\bea
H_\phys &=& -p_N = 6 \M^2 \sqrt{\frac{H_\phi a_0^3\mathcal{V}_0\,e^{3N}}{3}} \nn \\
&=& 6\M^3\, a_0^3\mathcal{V}_0\, e^{3N} H,
\eea
where the second equality arises from solving the Hamiltonian constraint for $p_N$, and in the last equality $H$ is the Hubble parameter defined as 
\begin{equation}
H = 
\frac{\altdot{a}}{\mathcal{N}a} = \frac{\dot{a}}{a}. 
\end{equation}
We use an overdot to denote the derivative with respect to proper time $t$, defined by $dt = \mathcal{N} d\tau$.  We can write the Hubble parameter in terms of the phase space variables of the constrained system using \eq{ham_cnstrnt}, 
\be \label{eq:Fried 1}
H^2 = \frac{H_\phi}{3\M^2a^3\mathcal{V}_0} \equiv \frac{\rho_\phi}{3\M^2},
\ee
where $\rho_\phi$ is the scalar field density. Other variables of observational interest can also be written in terms of the phase space variables (refer to Section \ref{EOM}).

\section{Polymer Quantization} \label{polymer}
We study a minimally coupled massive polymer quantized scalar field on a classical spatially flat FRW background. We begin by introducing the non-canonical variables $\Pi_\phi$ and $U_{\lambda}$, with
\be
U_\lambda = e^{i\lambda \phi}, \quad \lambda = M_{\lambda}^{-1},
\ee
where $M_{\lambda}$ is a fixed mass scale. These variables satisfy the Poisson algebra
\be
\label{PB}
\{ U_\lambda,  \Pi \} = i \lambda U_\lambda.
\ee
Polymer quantization proceeds by promoting the Poisson algebra of these variables to a commutator algebra on a Hilbert space with basis $\{\, \ket{\pp} \mid \pp \in \mathbb{R} \,\}$ with the inner product
\be
\braket{\pp}{\pp'} = \delta_{\pp,\pp'},
\ee
where $\delta$ is the generalization of the Kronecker delta to real numbers.
The basic operators $\hat{\Pi}$ and $\hat{U}_\lambda$ act as
\be
\hat{\Pi}\ket{\pp} = \pp \ket{\pp}, \quad \hat{U}_\lambda \ket{\pp} = \ket{\pp + \lambda},
\ee
where $\ket{\pp}$ is an eigenstate of the momentum operator $\hat{\Pi}$  and the eigenvalue $\pp$ has dimensions $(\text{mass})^{-1}$. 
The $\hat{U}_\lambda$ operator generates momentum translations. In this formalism, the field operator cannot be defined using the derivative ${\p \hat{U}_\lambda}/{\p \lambda}\big|_{\lambda=0}$, since $\hat{U}_\lambda$ is not weakly continuous in $\lambda$. An alternative definition of the field operator is
\be
\label{phi}
\hat{\phi} = -\frac{i}{2\lambda}( \hat{U}_\lambda - \hat{U}_\lambda^\dagger ).
\ee
In the limit $\lambda \rightarrow 0$, this can be solved to give the Schr\"odinger formula $\hat{U}_{\lambda} = e^{i\lambda \hat\phi}$.  Hence, we call $\lambda \rightarrow 0$ (or equivalently $M_\lambda \rightarrow \infty$) the Schr\"odinger limit.

We now use these operator definitions to calculate the expectation value of the scalar energy density $\rho_{\phi}$ (and hence the physical Hamiltonian) in a quantum state given by
\begin{equation}\label{eq:state}
	\ket{\psi} = \sqrt{\lambda} \sum_{k=-\infty}^{\infty} c_k \ket{\pp_k}, \quad \pp_{k} = k\lambda, \quad k \in \mathbb{Z}.
\end{equation}
We assume that the state is strongly peaked on the field momentum $\bar\pp$.  This can be achieved by sampling the expansion coefficients from a Gaussian distribution of width $\sigma$
\begin{equation}\label{eq:gaussian}
	c_k = \frac{1}{\pi^{1/4}\sigma^{1/2}} \exp  \l[ \frac{-(\pp_k - \bar\pp)^2}{2\sigma^{2}} \r] \exp ( - i \pp_k \phi_0 ).
\end{equation}
Here, $\phi_{0}$ is a $c$-number that we interpret below.

To calculate inner products and expectation values, we approximate sums by integrals
\begin{equation}
	\lambda \sum_{k=-\infty}^{\infty} \,\, \mapsto \,\, \int_{-\infty}^{\infty} d\pp_{k}.
\end{equation}
Under this prescription we have that (\ref{eq:gaussian}) implies that $\langle \psi | \psi \rangle = 1$,  and we obtain the following expectation values 
\begin{equation}
	\langle \Pi_{\phi} \rangle = \bar\pp, \,\,\, \langle U_{\lambda} \rangle = e^{i\Theta} e^{-\Sigma^{2}/4}, \,\,\, \langle\Pi_{\phi}^{2}\rangle = \bar\pp^{2} + \frac{\sigma^{2}}{2},
\end{equation}
where
\begin{equation}
	\Theta = \frac{\phi_{0}}{M_{\lambda}}, \quad \Sigma = \frac{1}{M_{\lambda}\sigma}.
\end{equation}
Using the operator definition (\ref{phi}), this then yields the expectation value of $\phi$:
\begin{equation}
	\langle \phi \rangle = M_{\lambda} e^{-\Sigma^{2}/4}\sin \Theta, \quad \lim_{M_{\lambda} \rightarrow \infty} \langle \phi \rangle = \phi_{0}.
\end{equation}
Hence, $\phi_{0}$ is interpreted as the expectation value of the field amplitude in the Schr\"odinger limit $M_{\lambda} \rightarrow \infty$.

Using these formulae, we can evaluate the expectation value of the scalar field density
\begin{multline}
\label{effrho}
 \langle \rho_\phi \rangle = \frac{\bar{\pp}^2+\frac{1}{2} \sigma^{2}}{2a^6\mathcal{V}_0^2} + \frac{m^2 M_{\lambda}^{2}}{4}\l[ 1-\cos(2\Theta) e^{-{\Sigma^2}} \r].
\end{multline}
We use this in (\ref{eq:Fried 1}) to write the semi-classical Friedmann equation and physical Hamiltonian
\begin{equation}
	H^{2} = \frac{\langle \rho_\phi \rangle}{3\M^2}, \quad H_\text{phys} = 3 \sqrt{3} \M^{2} a^{3} \langle \rho_\phi \rangle^{1/2},
\end{equation}
in the e-fold time gauge.  This follows from taking the expectation value of the Hamiltonian constraint (\ref{ham_cnstrnt}) with geometric variables treated classically; i.e., as $c$-numbers.  We will study the dynamics generated by these expression in the next section.

Finally, we note that at no point in the above calculations were we obliged to take $\sigma$ to be a constant.  It is plausible that the width of the semi-classical state describing $\phi$ varies in time; hence, we will also study situations where the width has a power-law dependence on the scale factor:
\begin{equation}
	\sigma = M_{\sigma}^{-1} (M_{\sigma}^{3} \mathcal{V}_{0} a^{3})^{l/2}.
\end{equation}
Here, $M_{\sigma}$ and $l$ are constants.  The factor in brackets was selected to ensure that $\sigma$ is invariant under the transformations (\ref{eq:transformations}).  We assume that $l \ge 0$ so that $\sigma$ does not diverge in the early time $a \rightarrow 0$ limit.  We also demand that $e^{-\Sigma^{2}} \rightarrow 1$ at late times, which means $l < 2$.  With these restrictions, the late time limit of the effective scalar density is
\begin{multline}\label{eq:rho at late times}
	 \langle \rho_\phi \rangle = \left[ \frac{\bar{\pp}^2}{2a^6\mathcal{V}_0^2} + \mathcal{O}(a^{3(l-2)}) \right]  \\ + \left[\frac{m^2 M_{\lambda}^{2}}{2} \sin^{2}\l(\frac{\phi_{0}}{M_{\lambda}}\r) + \mathcal{O}(a^{-3l}) \right].
\end{multline}
The first term in square brackets is essentially the scalar field kinetic energy with a quantum width correction $\mathcal{O}(a^{3(l-2)})$.  The second term gives the effective scalar field potential at late times, and we see that it matches the natural inflation potential (\ref{NI}) if the $\mathcal{O}(a^{-3l})$ term is neglected.  We note that if $l=0$, the $\mathcal{O}(a^{-3l})$ behaves like the cosmological constant term in the HNI potential (\ref{eq:HNI Potential}); we discuss this in greater detail in section \ref{sec:HNI} below.

In addition to the late time limit, one can examine the behaviour of $\langle \rho_{\phi} \rangle$ as $M_{\lambda} \rightarrow \infty$.  We obtain:
\begin{equation}\label{eq:density limit}
	\langle \rho_{\phi} \rangle \rightarrow \frac{1}{2} \langle \dot\phi^{2} \rangle + \frac{1}{2} m^{2} \langle \phi^{2} \rangle_\text{S}.
\end{equation}
Here, the expectation value of $\dot\phi^{2}$ is
\begin{equation}
	\langle \dot\phi^{2} \rangle = \frac{\langle \Pi_{\phi}^{2} \rangle}{\mathcal{V}_{0}^{2}a^{6}} = \dot{\phi}_{0}^{2} + \frac{\Msig^{3l-2}a^{3(l-2)}\mathcal{V}_0^{l-2}}{2},
\end{equation}
and $\langle \phi^{2} \rangle_\text{S}$ represents the expectation value of $\phi$ as if we interpreted $c_{k}$ as a momentum wavefunction in Schr\"odinger quantum mechanics:
\begin{equation}
	\langle \phi^{2} \rangle_\text{S} = \int_{\infty}^{\infty} dp_{k} \, c_{k}^{*} \left( i \frac{\di}{\di p_{k}} \right)^{2} c_{k} = \phi_{0}^{2} + \frac{M_{\sigma}^{2-3l}}{2(\mathcal{V}_{0}a^{3})^{l}}. 
\end{equation}
If we neglect quantum uncertainty effects,
\begin{equation}
\langle \dot\phi^{2} \rangle \approx \dot{\phi_0}^{2}, \quad \langle \phi^{2} \rangle_\text{S} \approx \phi_{0}^{2},
\end{equation}
then (\ref{eq:density limit}) reduces to the usual scalar field density for chaotic inflation.

\section{Semiclassical dynamics} \label{semiclassical}
\subsection{Equations of Motion} \label{EOM}
The semiclassical equations of motion are given by Hamilton's equations
\be
\l( \frac{d\phi_0}{dN}, \frac{d\bar{\pp}}{dN} \r) = \l( \frac{\p H_\phys}{\p \bar{\pp}}, -\frac{\p H_\phys}{\p \phi_0} \r).
\ee
It is useful to change variables to the proper time $t$:
\begin{equation}
	\frac{d}{dN} = \frac{a}{\dot{a}} \frac{d}{dt} = \frac{1}{H} \frac{d}{dt}. 
\end{equation}
The equations of motion and Friedmann equation are explcitly
\begin{subequations}
\begin{align}\label{EOM 2 a}
\frac{d\phi_0}{dt} & = \frac{\bar{\pp}}{\mathcal{V}_{0} a^{3}} , \\ \label{EOM 2 b}
\frac{d\bar{\pp}}{dt} & = -\frac{\mathcal{V}_{0} a^{3} M_{\lambda} m^2  e^{-\Sigma^2} \sin(2\Theta) }{ 2 }, \\
H^{2} & =  \frac{1}{3\M^{2}} \left[ \frac{1}{2} \langle \dot\phi^{2} \rangle + V_\eff(\phi_{0},a) \right]. \label{eq:Friedmann}
\end{align}
\end{subequations}
Here, the effective potential is given by
\begin{equation}
	V_{\eff}(\phi_{0},a) = \frac{m^{2}M_{\lambda}^{2}}{4} \left[ 1 - \exp\left( -\frac{M_{\sigma}^{2-3l}}{M_{\lambda}^{2}\mathcal{V}_{0}^{l} a^{3l}} \right) \cos \frac{2\phi_{0}}{M_{\lambda}}   \right],
\end{equation}
and the expectation value of $\dot\phi^{2}$ is
\begin{equation}
	\langle \dot\phi^{2} \rangle = \frac{\langle \Pi_{\phi}^{2} \rangle}{\mathcal{V}_{0}^{2}a^{6}} = \dot{\phi}_{0}^{2} + \frac{\Msig^{3l-2}a^{3(l-2)}\mathcal{V}_0^{l-2}}{2}.
\end{equation}
Finally, we note that equations (\ref{EOM 2 a}) and  (\ref{EOM 2 b}) can be combined into a single second order equation:
\begin{equation}
	\ddot{\phi}_{0} + 3 H \dot{\phi}_{0} + \partial_{\phi_{0}} V_{\eff}(\phi_{0},a) = 0.
\end{equation}

\subsection{Dynamics}

We can re-write the equations of motion as an autonomous dynamical system in order to extract its qualitative features. We demand that the quantum width correction to the kinetic energy be smaller than the correction to the potential in (\ref{eq:rho at late times}), which implies $l<2$.

We introduce dimensionless quantities: 
\begin{subequations}
\begin{align}
	\Phi & = 2\phi_{0}/M_{\lambda}, \\ 
	X & = \exp \l( -M_\text{Pl}^{-2} m^{-1} a^{-3} \mathcal{V}_0^{-1} \r), \\ 
	Y & = \arctan(\Phi'), \\ 
	Z & = \exp\left( -{M_{\sigma}^{2-3l}}{M_{\lambda}^{-2}\mathcal{V}_{0}^{-l} a^{-3l}} \right),\\   
	\tau & =\sqrt{2}mt, \\ 
	\beta & = \sqrt{3} M_{\lambda}/(2\M),
\end{align}
\end{subequations}
in terms of which the equations of motion can be written as an autonomous dynamical system:
\begin{subequations}\label{eq:dynamical system}
\begin{eqnarray}
	 \Phi' & = & \tan (Y),  \\
	 X' &=& -\frac{\beta}{\sqrt{2}}X \ln(X) \sec(Y) \,\mathscr{H}(\Phi,X,Y,Z), \\
	 Y' & = & -\frac{\beta}{\sqrt{2}}\sin (Y) \,\mathscr{H}(\Phi,X,Y,Z) \nn \\ & & -\frac{1}{2}Z\sin(\Phi) \cos^{2}(Y),  \\ 
	Z' & = & -\frac{\beta l}{\sqrt{2}} Z \ln (Z) \sec (Y) \,\mathscr{H}(\Phi,X,Y,Z).	
\end{eqnarray}
\end{subequations}
with 
\begin{equation}
\mathscr{H}= \sqrt{1 -\cos^{2}(Y)\l\{Z\cos(\Phi)-\l[ \frac{3 \ln (X)}{4\beta^2 \ln (Z)} \r]^2 \r\}}. \nn 
\end{equation}
Here, we use a prime $'$ to denote differentiation with respect to $\tau$.  We have that $Y \in (-\pi/2,\pi/2)$ and $Z,X \in (0,1]$.  Note that the system is invariant under $\Phi \mapsto \Phi \pm 2\pi$.  This means we can choose $\Phi \in [-\pi,\pi)$, which yields a compact phase space.

One can recover the dynamics of the scale factor in certain cases by examining the so-called ``Hubble slow-roll'' parameter
\begin{equation}
	\epsilon_\text{H} = - \frac{\dot{H}}{H^{2}}, \quad \ddot{a} = H^{2}a (1 - \epsilon_\text{H}).
\end{equation}
It follows that when $\epsilon_\text{H} < 1$, the cosmological expansion is accelerating $\ddot{a}>0$.  In terms of the dynamical variables defined above, we have
\begin{eqnarray}
\label{epsilon_H}
	\epsilon_\text{H} &=& \frac{24\beta^4\ln(Z)^2[lZ\ln(Z)\cos(\Phi)-2\tan^2(Y)] }{16\beta^4\ln(Z)^2[Z-\sec^2(Y)] - 9 \ln(X)^2} \nn \\
	&+&\frac{ 9(l-1)\ln(X)^2}{16\beta^4\ln(Z)^2[Z-\sec^2(Y)] - 9 \ln(X)^2}.
\end{eqnarray}
In the early time limit we have,
\begin{equation}
\lim_{\tau \rightarrow -\infty} X = 0, \quad \lim_{\tau \rightarrow -\infty} Y= \pm \frac{\pi}{2}, \quad \lim_{\tau \rightarrow -\infty} Z  = 0.
\end{equation}
In this limit equation (\ref{epsilon_H}) gives
\begin{equation}
\epsilon_\text{H} \sim 3 \quad \Rightarrow \quad a \propto (\tau-\tau_0)^{1/3}.
\end{equation}
This is the expected scale factor behaviour for a massless scalar field, and we see that $a \rightarrow 0$ as $\tau \rightarrow \tau_{0}^{+}$.  That is, the big bang singularity is a past attractor of the system for all parameters.  In other words, the polymer quantization scheme employed in this paper does nothing to avoid the classical big bang singularity.

In order to study the inflationary phase and late time behaviour of the dynamics, it is useful to consider the simplifying approximation that quantum corrections to the kinetic energy term in the Friedman equation (\ref{eq:Friedmann}) are negligible:
\begin{equation}
	\langle \dot\phi^{2} \rangle \approx \dot{\phi}_0^{2}.
\end{equation}
The choice $l \in [0,2)$ ensures this approximation is valid at late times. The dynamical system then reduces to 
\begin{subequations}\label{eq:dynamical system}
\begin{eqnarray}
	 \Phi' & = & \tan (Y), \label{eq:Phi prime} \\
	 \label{eq:Y prime} Y' & = & -\frac{\beta}{\sqrt{2}}\sin (Y) \sqrt{1 -Z\cos^{2}(Y) \cos(\Phi)}  \nn \\ & & -\frac{1}{2}Z\sin(\Phi) \cos^{2}(Y),  \\ 
	Z' & = & -\frac{\beta l}{\sqrt{2}} Z \ln (Z) \sec (Y) \nn \\ & & \times \sqrt{1 - Z\cos^{2}(Y)\cos(\Phi)}  . \label{eq:Z prime}
	\end{eqnarray}
\end{subequations}

The system's trajectory through phase space is qualitatively different for the $l=0$ and $l \in (0,2)$ cases.

When $l\neq 0$, equation (\ref{eq:Z prime}) implies that $Z=0$ and $Z=1$ are repulsive and attractive invariant submanifolds for the system (\ref{eq:dynamical system}). That is,
\begin{equation}
\lim_{a\rightarrow -\infty} Z = 0, \quad \lim_{a \rightarrow +\infty} Z = 1.
\end{equation}
Any trajectory accumulates on the $Z=1$ surface since $Z' \geq 0$ for all initial data. Near the $Z=1$ submanifold, the dependence of  the dynamics  on $l$ is trivial and slow roll inflation is recovered.

In the late time limit ($a \rightarrow \infty$ and $Z \rightarrow 1$), equation (\ref{epsilon_H}) gives
\begin{equation}
a \sim \tau^{2/3l}, \quad \epsilon_\text{H} = \tfrac{3}{2}l.
\end{equation}
If $l < 2/3$, we see that the late time phase is accelerating and as $l \rightarrow 0$ we recover a late time de Sitter phase with exponential expansion.  

\subsection{Special case: $l = 0$}\label{sec:HNI}

\begin{figure}
	\includegraphics[width=\columnwidth]{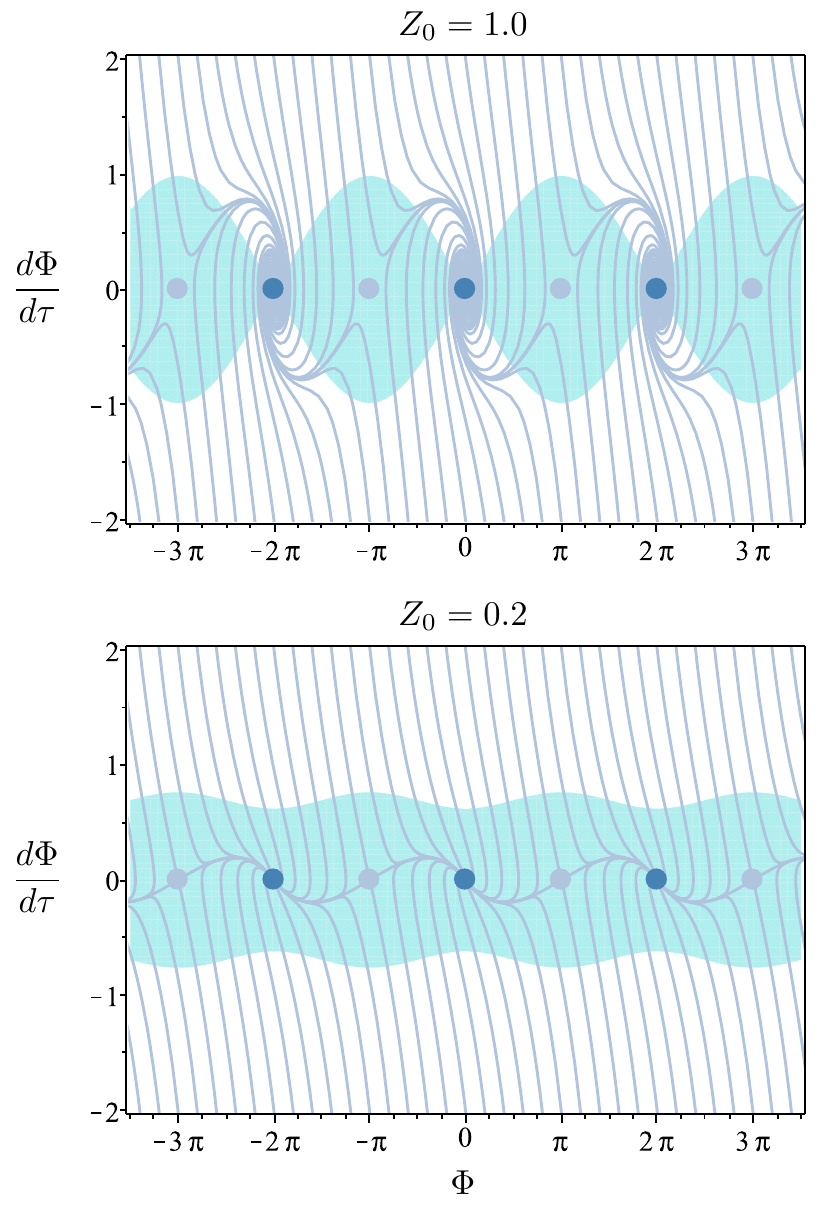}
	\caption{Phase portraits for the $l = 0$ case.  We have selected $\beta=1.0$. The dark circles are the attractive fixed points $\Gamma_{+}$ and the light circles are the repulsive fixed points $\Gamma_{-}$.  Note that the periodicity of the orbits in the $\Phi$ direction.  The shaded regions indicate when the condition for inflation $\epsilon_\text{H}<1$ is satisfied.}\label{fig:flat phase}
\end{figure}

\begin{figure}
	\includegraphics[width=\columnwidth]{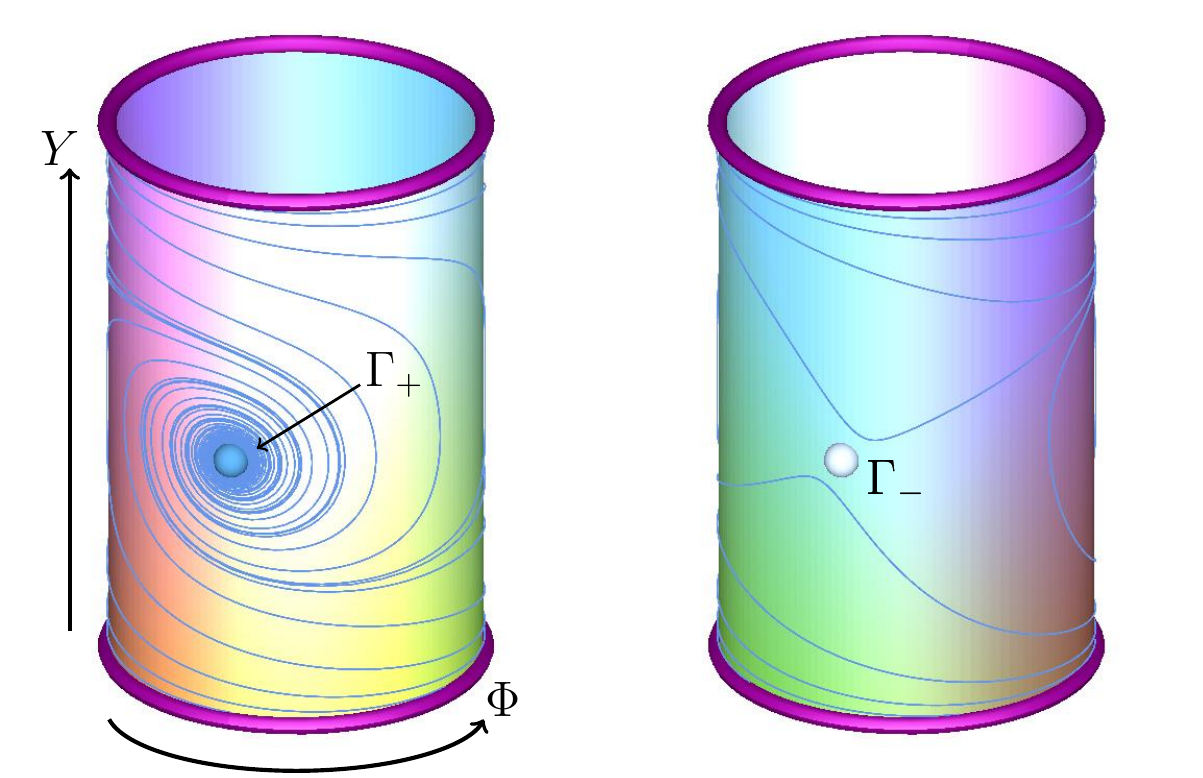}
	\caption{Phase portraits for the $l = 0$ case drawn on a cylinder.  We have selected $\beta=1.0$ and $Z_{0} = 1.0$.}\label{fig:cylinder phase}
\end{figure}

We now consider the $l=0$ case in more detail.  This assumption implies that the semi-classical state (\ref{eq:state}) is not explicitly dependent on $a$.   When $l=0$, equation (\ref{eq:Z prime}) implies that
\begin{equation}
	Z = \text{constant} = e^{-M_{\sigma}^{2}/M_{\lambda}^{2}} \equiv Z_{0}.
\end{equation}
In this case, the dynamical system is essentially comprised of (\ref{eq:Phi prime}) and (\ref{eq:Y prime}) with $Z_{0}\in (0,1)$ as a parameter.  In figures \ref{fig:flat phase} and \ref{fig:cylinder phase} we plot phase portraits of the $l = 0$ system for various parameters.  The shaded regions in each panel indicate where the condition for inflation ($\epsilon_\text{H}<1$) is satisfied.
\begin{table}
\begin{ruledtabular}
\begin{tabular}{ccccc}
fixed pt.\ & $\Phi$  & $Y$ & $Z_{0}$ & classification \\ \hline
$\Gamma_{+}$ & $2n\pi$  & 0 & $\in(\beta^{2}/(4+\beta^{2}),1]$ & spiral attractor \\
 & & &  $\in(0,\beta^{2}/(4+\beta^{2})]$ & attractor \\ 
$\Gamma_{-}$ & $(2n+1)\pi$  & 0 & $\in(0,1]$ & saddle 
\end{tabular}
\end{ruledtabular}
\caption{Fixed points of the dynamical system (\ref{eq:dynamical system}) when $l = 0$.}\label{eq:fixed points}
\end{table}

The fixed points of the $l=0$ dimensionally reduced dynamical system are given in Table \ref{eq:fixed points}, along with their classification.  As discussed above, at early times the system will behave as a massless scalar field cosmology near the big bang singularity in general relativity.  On the other hand, if we assume $Z_{0} \in (0,1)$ we have that near the future attractor $\Gamma_{+}$
\begin{equation}
\label{late time inflation}
	\epsilon_\text{H} \sim 0 \quad \Rightarrow \quad a \propto e^{c\tau},
\end{equation}
where $c$ is a constant.  Hence, at late times the universe exhibits de Sitter like acceleration.  To understand this late time behaviour, we can examine the effective potential when $l =0$:
\begin{equation}\label{eq:l=0 potential}
	V_{\eff}(\phi,a) = \Lambda^4 \l[ 1 - \cos(\frac{\phi}{ f }) \r] + \rho_\text{v}.
\end{equation}
This is identical to the HNI inflation potential (\ref{eq:HNI Potential}) with mass parameters related to $m$, $M_{\lambda}$ and $M_{\sigma}$ by:
\begin{gather}\nn
	f = \frac{M_{\lambda}}{2}, \quad \Lambda^{4} = \frac{m^{2}M_{\lambda}^{2}e^{-M_{\sigma}^{2}/M_{\lambda}^{2}}}{4}, \\ \rho_\text{v} =  \frac{m^{2}M_{\lambda}^{2}(1-e^{-M_{\sigma}^{2}/M_{\lambda}^{2}})}{4}.
\end{gather}
Note that $\rho_\text{v}$ is essentially a cosmological constant term that dominates at late times.  These formulae may be inverted to yield
\begin{gather}\nn
	m^{2} = \frac{\Lambda^{4}+\rho_\text{v}}{f^{2}}, \quad M_{\lambda} = 2f, \\ M_{\sigma}^{2} = 4f^{2}\ln \left( 1 + \frac{\rho_\text{v}}{\Lambda^{4}} \right).
 \end{gather}
If the vacuum contribution to $V_{\eff}$ is negligible ($\rho_\text{v} \ll \Lambda^{4}$), then the potential is basically that of natural inflation (\ref{NI}) and we can use Planck results \cite{Planck} to constrain $M_{\lambda}$ to be super-Planckian:
\begin{equation}
 M_{\lambda} \gtrsim 3.5\,\M.
\end{equation}

It is, of course, somewhat tempting to associate the vacuum contribution with the current observed value of the cosmological constant via $\Lambda_\text{obs} = \rho_\text{v}/\M^{2} \sim 10^{-31}\,\M$.  But this leads to an extremely small value of $M_{\sigma}$ if we take reasonable values for $f$ and $\Lambda$:
\begin{align}\nn
	\frac{M_{\sigma}}{\M} & = 2 \times 10^{-12} \left( \frac{f}{\M} \right) \left( \frac{\Lambda}{10^{-3} \, \M} \right)^{-4} \left( \frac{\Lambda_\text{obs}}{\M^{2}} \right)^{1/2} \\ & \sim 2 \times 10^{-43} \left( \frac{f}{\M} \right) \left( \frac{\Lambda}{10^{-3} \, \M} \right)^{-4}.
\end{align}
In other words, if one wishes to explain dark energy by using the framework presented here, $M_{\sigma}$ would have to be tuned to a very small value.

\begin{figure}
	\includegraphics[width=\columnwidth]{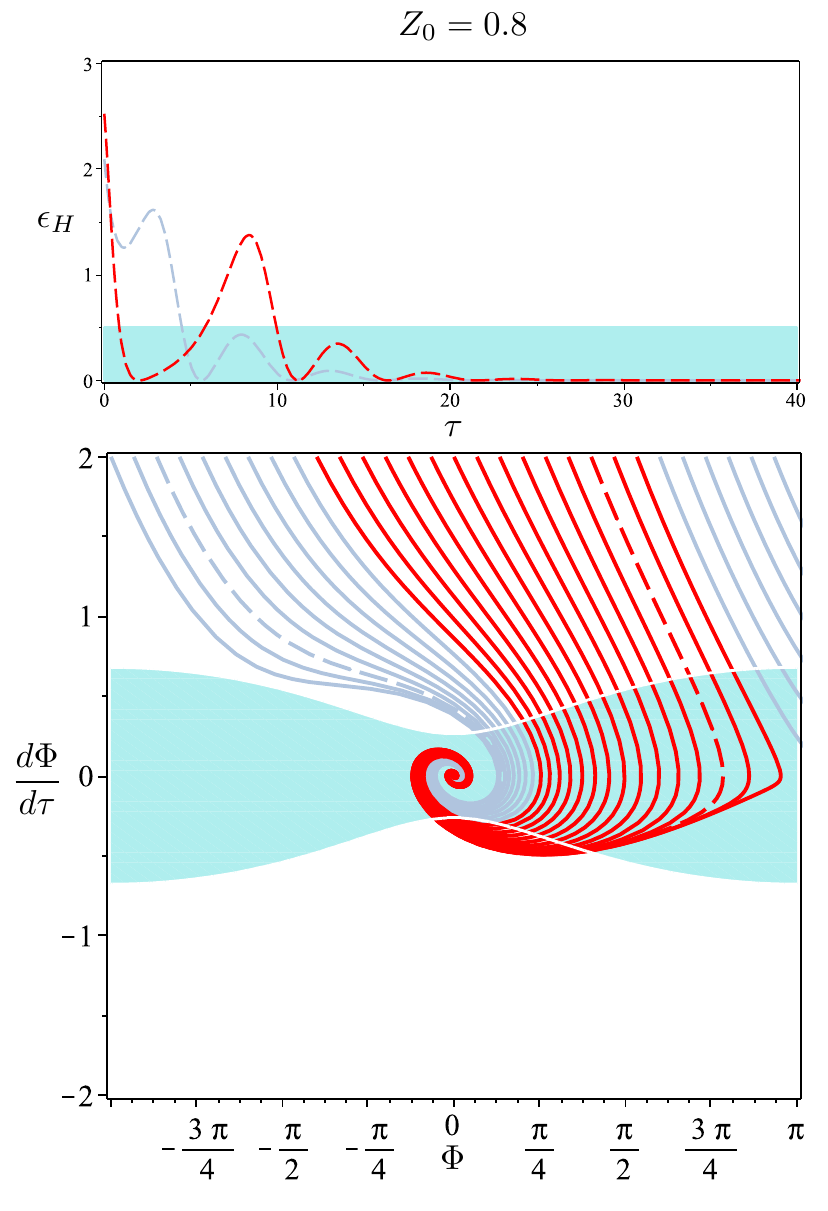}
	\caption{Time evolution of the $\epsilon_\text{H}$ slow roll parameter (\emph{top}) and phase portrait (\emph{bottom}) for the $(l,\beta,Z_{0})=(0,1,0.8)$. The shaded regions indicate when the condition for ``slow-roll'' inflation is satisfied: $\epsilon_\text{H} < 0.5$. Trajectories coloured in red exit the slow roll inflationary phase at finite time and later re-enter.}\label{fig:efolds phase}
\end{figure}

Finally, we note that even though the potentials (\ref{eq:HNI Potential}) and (\ref{eq:l=0 potential}) are algebraically indistinguishable, the polymer quantized model we have written down here is different from HNI in at least one important respect:  In HNI, there exists a ``waterfall field'' that serves to terminate inflation after a finite time.  In the current model no such field is present, which means that de Sitter space is a future attractor of the system and inflation technically never ends.  In addition, there are cosmological trajectories with two distinct  inflationary epochs, the first of which is finite duration.  This is illustrated in figure \ref{fig:efolds phase}, where we show the time evolution of $\epsilon_\text{H}$ and a phase portrait of the system for the $Z_{0} = 0.8$ case ($\rho_\text{v} \not\approx 0$).  In both panels, the shaded region represents the portion of the phase plane associated with $\epsilon_\text{H} < 0.5$, which we take to be the operational definition of slow-roll inflation for the purpose of the current discussion.  In the top panel, we show $\epsilon_\text{H}$ versus $\tau$ for a trajectory which enters, exits and re-enters the slow-roll regime (the red dashed line), as well as a trajectory that enters slow-roll once and never leaves (the blue dashed line).  Similarly, the trajectories which enter and then exit the slow roll regions are coloured red and trajectories which cross into the slow roll region only once are coloured blue in the bottom panel.  
\begin{figure}
\includegraphics[width = \columnwidth]{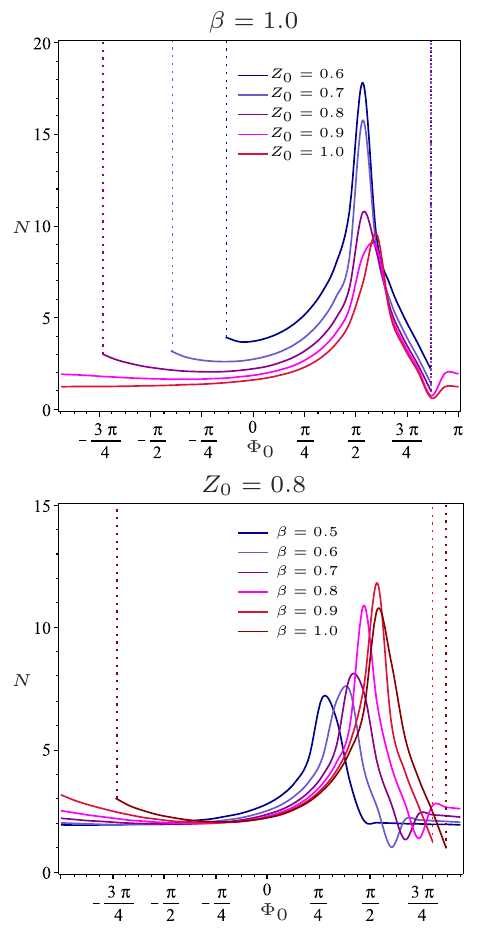}
\caption{These figures show how the duration of the initial slow-roll inflationary phase (as measured in e-folds $N$) varies with $\beta$, $Z_0$, and the initial field amplitude $\Phi_{0}$ when $l = 0$.  We chose initial data with $d\Phi/d\tau = 2$.  Note that some initial data leads to an infinite amount of initial inflation, as indicated by the dotted lines.  The trajectories with finite $N$ are coloured red in figure \ref{fig:efolds phase}. } \label{fig:number of efolds}.
\end{figure}

The red trajectories of figure \ref{fig:efolds phase} can be interpreted as cosmologies that undergo a phase of slow roll inflation with definite start and end before a secondary inflationary phase.  On the other hand, once slow roll inflation starts for the blue trajectories it continues indefinitely.  That is, the number of e-folds of accelerated expansion in the first inflationary epoch is finite (infinite) for red (blue) trajectories.  It is possible to numerically determine the duration of the first inflation phase (if it is finite) for various choices of initial data, $\beta$ and $Z_{0}$.  In figure \ref{fig:number of efolds}, we present the number of e-folds in the initial inflationary phases as a function of the initial field amplitude $\Phi|_{\tau =0} = \Phi_{0}$ if $d\Phi/d\tau|_{\tau = 0} = 2$.

\section{Discussion}

We studied the semiclassical dynamics of a polymer quantized scalar field on a spatially flat Friedman-Robertson-Walker background. The semiclassical state was chosen to be a fairly general dilation invariant coherent state.  Our objective was to study this model with a different polymer quantization scheme than used in \cite{Hassan:2014sja}.  The prescription followed here results in an effective inflationary potential resembling that of natural inflation. The early time generic ``polymer" inflation phase obtained using the other prescription is absent, and the semiclassical dynamics exhibits a big bang singularity.

When the semiclassical state has no explicit scale factor (time) dependence the effective potential is exactly that of hybrid natural inflation. This implies that sufficient inflation can be obtained even for sub-Planckian values of the polymer scale, and, in contrast to the original natural inflation model, the tensor-scalar ratio may be consistent with observations. However, unlike the hybrid natural inflation model we have no auxiliary field to end inflation; i.e. for most choices of parameters inflation is eternal. One could, of course, consider generalizations of the current model involving a waterfall mechanism to end inflation similar to that employed in \cite{Ross:2009hg,Ross:2010fg,Carrillo-Gonzalez:2014tia,Vazquez:2014uca,German:2016umn,Ross:2016hyb,German:2017pfu}.  Both a complete analysis of the range of parameter space consistent with observations and generalizations of this model involving additional fields are an interesting avenue for further work.

In this article, we have only considered the polymer quantization of a homogeneous scalar field and focused on the resulting natural inflation like semiclassical dynamics. In order to develop a complete and consistent theoretical model for studying polymer effects in the CMB anisotropies and large scale structure, quantum inhomogeneities in both the gravity and matter sectors need to be considered. Thus we need to go beyond polymer quantum mechanics to polymer quantum field theory, which is complicated by the necessity of defining a suitable gradient operator for the polymer quantized field.  One possibility is to follow a procedure similar to that outlined in \cite{husain:2010}.  Furthermore, since the gauge invariant variables commonly used to describe cosmological perturbations mix geometric and matter degrees of freedom, it would be interesting to consider the simultaneous polymer quantization of both gravity and matter sectors. These are also directions for future work.

\begin{acknowledgments}

We would like to thank Edward Wilson-Ewing for comments on this manuscript. We are supported by NSERC of Canada.  In addition, this research was supported in part by Perimeter Institute for Theoretical Physics. Research at Perimeter Institute is supported by the Government of Canada through Innovation, Science and Economic Development Canada and by the Province of Ontario through the Ministry of Research, Innovation and Science.

\end{acknowledgments}

\vfill

 \bibliography{polymer}
\end{document}